\documentclass[12pt]{article}
\usepackage{amssymb,latexsym,amsmath}
\newcommand{\ft}[2]{{\textstyle\frac{#1}{#2}}}
\def\Re{\mathop{\rm Re}\nolimits}
\def\Im{\mathop{\rm Im}\nolimits}
\def\N{\mathcal N}

\def\rme{{\rm e}}

\def\rmd{{\rm d}}

\def\veps{{\varepsilon}}
\newsavebox{\uuunit}
\sbox{\uuunit}
    {\setlength{\unitlength}{0.825em}
     \begin{picture}(0.6,0.7)
        \thinlines
        \put(0,0){\line(1,0){0.5}}
        \put(0.15,0){\line(0,1){0.7}}
        \put(0.35,0){\line(0,1){0.8}}
       \multiput(0.3,0.8)(-0.04,-0.02){12}{\rule{0.5pt}{0.5pt}}
     \end {picture}}
\newcommand {\unity}{\mathord{\!\usebox{\uuunit}}}

\csname @addtoreset\endcsname{equation}{section}

   \usepackage{cite}

\begin{document}

\begin{titlepage}
\begin{flushright}
KUL-TF-07/13\\
arXiv:0708.2829 [hep-th]
\end{flushright}
\vspace{.5cm}
\begin{center}
\baselineskip=16pt
{\LARGE Effective action for the field equations\\ \vskip 0.2cm
of charged black holes
}\\
\vfill
{\large Antoine Van Proeyen, Bert Vercnocke 
  } \\
\vfill
{\small  Instituut voor Theoretische Fysica, Katholieke Universiteit Leuven,\\
       Celestijnenlaan 200D B-3001 Leuven, Belgium.
      \\[2mm] }
\end{center}
\vfill
\begin{center}
{\bf Abstract}
\end{center}
{\small We consistently reduce the equations of motion for the bosonic
$N=2$ supergravity action, using a multi-centered black hole ansatz for
the metric. This reduction is done in a general, non-supersymmetric
setup, in which we extend concepts of BPS black hole technology. First
we obtain a more general form of the black hole potential, as part of an
effective action for both the scalars and the vectors in the supergravity
theory. Furthermore, we show that there are extra constraints specifying
the solution, which we calculate explicitly. In the literature, these
constraints have already been studied in the one-center case. We also
show that the effective action we obtain for non-static metrics, can be
linked to the ``entropy function'' for the spherically symmetric case, as
defined by Sen and Cardoso et al.}\vspace{2mm} \vfill \hrule width 3.cm
{\footnotesize \noindent e-mails: antoine.vanproeyen@fys.kuleuven.be,
bert.vercnocke@fys.kuleuven.be }
\end{titlepage}
\addtocounter{page}{1}
 \tableofcontents{}
\newpage
\section{Introduction}\label{sect-intro}
It has been more than 10 years now since the attractor mechanism in
supersymmetry was discovered, in the context of $N=2$ extremal black
holes \cite{Ferrara:1995ih, Strominger:1996kf,Ferrara:1996dd}. Since
then, it has been realized that the attractor phenomenon is not
restricted to the supersymmetry of the solution, but rather is a property
related to the extremality of black holes. This was first proposed in
\cite{Ferrara:1997tw}, where also the concept of a black hole potential
was introduced, and studied in detail later in
\cite{Tripathy:2005qp,Goldstein:2005hq,Kallosh:2006bt}. In the latter,
the non-extremal attractor equation was derived in great detail. All this
work was done for static black holes.

In the context of string theory, the attractor phenomenon has also been
well studied for BPS black hole systems \cite{Moore:1998pn}. In
particular, it was shown \cite{Denef:2000nb, Bates:2003vx} that solutions
exist with multiple centers. They can be described in the low energy
limit as stable configurations of several coupled black holes. These no longer have to be described by a static metric.

We do not have the intention of giving a detailed description of the
attractor phenomenon. We do, however, want to extend some of the ideas
concerning the reduced action one generically writes down for such
extremal black holes, which is the basis of a detailed study of the
attractor equations. In the case of a static metric, a spherically
symmetric metric ansatz is generically proposed -- see for instance
\cite{Ferrara:1997tw,Kallosh:2006bt}. Using this ansatz, one eliminates
the vectors through their equation of motion and integrates out angular
dependence of the supergravity action. In this way one obtains a reduced
action, governing the dynamics of the scalars, introducing a
charge-dependent black hole potential along the way. For completeness, a
(Hamiltonian) constraint has to be imposed on the system. Another method
is based on integrating out angular dependence and a Legendre transform
of the reduced Lagrangian is carried out. The stationary points of the
entropy function thus obtained \cite{Sen:2005wa,Cardoso:2006xz} give the
attractor equations. When one studies multicenter solutions, as is done
in \cite{Denef:2000nb}, one does not start from the standard form of the
supergravity action exhibiting manifest four-dimensional covariance.
Instead, one can derive the attractor behavior from a four-dimensional
electric/magnetic duality invariant action.  However, the author has not
checked compatibility of this action with the full four-dimensional
metric. When doing so, by comparing with the four-dimensional Einstein
equations, constraints may have to be imposed on the system. In
\cite{Denef:2000nb}, this is avoided by immediately imposing BPS
conditions on the reduced action. The BPS solutions are then supposed to
satisfy these extra constraints.

We would like to obtain such an action in another way. We do this by
studying the information contained in the Einstein equations. The
Einstein equations, together with the equations of motion for the scalars
and those of the vectors, govern the dynamics of the black hole solution.
Knowing this, we will try to integrate (some of these) Einstein equations
to give an effective action. The Einstein equations that cannot be
derived from this effective action, will become constraints that have to
be imposed on the solution. We will show that the effective action
correctly describes the scalar and metric dynamics as well as those of
the vectors. In fact, the effective action we obtain in this way, is
exactly the one that has already been studied \cite{Denef:2000nb}, but
now we also arrive at extra constraints that were not found then. We also
show that the action we obtain is analogous to the concept of the entropy
function of Sen \cite{Sen:2005wa} and Cardoso et al.
\cite{Cardoso:2006xz}, in the sense that our effective action is (minus)
the entropy function for spherically symmetric metrics.

We apply our method to stationary spacetimes. Also, all the fields will
be required to be time-independent. Both the static and non-static cases
are discussed, in a non-supersymmetric setup. This will allow us to
propose an effective action, first  for static but not necessarily
spherically symmetric metrics. When we do look at spherical symmetry, we
see that our effective action is the same as the reduced action in the
literature. We also obtain the same constraint. However, the method we
use is different from a dimensional reduction. There is a subtlety in the
choice of independent variables. We will come back to this later. When we
apply our machinery to non-static black hole systems, we arrive again at
an effective action for the scalars, together with extra constraints.
These constraints have not been found before.  Also, we can
straightforwardly generalize the concept of a black hole potential.

The outline of this paper is as follows. We begin by elaborating on
results for a static black hole metric ansatz, in section
\ref{sect-static}. This will be more than just a review of what is known
already for extremal black holes. We will bring our method into practice
for a generic static metric ansatz, to establish the key features of our
consistent reduction of the supergravity action by means of the field
equations. This will lead us to proposing an effective action and a
dynamical constraint restricting the solution. Where the original
supergravity action had the scalars and the (four-dimensional) vectors as
independent variables, this will no longer be the case here. The
effective action reproduces the vector and scalar field equations, but
only after a specific reparametrization of the vector contribution, in
which we switch to new three-dimensional vector variables. Afterward, we
make a specific choice of the metric and obtain indeed the same results
as in the literature, including the notion of a black hole potential.

In section \ref{sect-nonstatic}, we apply the same steps toward an
effective action, this time for a general non-static metric ansatz
admitting multicenter black holes. Again, the four-dimensional vectors
will no longer be the independent variables in the effective action we
obtain, but  new three-dimensional ones will have to be considered.  We
will arrive at a generalization of the concept of a black hole potential.
Furthermore, there will also be constraints generalizing those that we
had for a static metric and we will present them in detail.

In section \ref{sect-comparing_actions}, we explain how the effective
action we obtain is different from the supergravity action we started
from. In fact, we start from a manifestly covariant action in four
dimensions, which is not invariant under electric/magnetic duality
rotations. The effective Lagrangian we obtain however, does transform as
a scalar under electric/magnetic duality. But because we switched to
three-dimensional gauge vectors as independent variables, it will no
longer be covariant with respect to the full four-dimensional spacetime.

Finally, we conclude in section \ref{sect-conclusion}.

\section{Introducing our method: static metrics\label{sect-static}}
Before applying our method to derive an effective action for multicenter
black holes, we will look to static metrics. First, this will allow us to
introduce the key ideas of our reduction of the equations of motion in a
simpler setup. Secondly, we will be able to link the solutions we find to
the results known in the literature. In particular, this will make it
clear how we generalize the concept of a black hole potential, both in
static configurations and later also in non-static ones.

A static spacetime metric\footnote{'Static' means that it admits a
global, nowhere zero, timelike hypersurface orthogonal Killing vector
field. A generalization are the 'stationary' spacetimes, which admit a
global, nowhere zero timelike Killing vector field. In that case the
components $g_{0m}$ could be nonzero. We will turn first our attention to
these static spacetimes.} has the general form:
\begin{eqnarray}
&&\rmd s^2 = -\rme^{2U}\rmd t^2 + \rme^{-2U}\gamma_{mn}\rmd x^m\rmd x^n\,,\label{stationaryMetric}\\
&&\mbox{where:}\quad\partial_t U = \partial_t\gamma = 0\,.\nonumber
\end{eqnarray}
In the following, $\gamma_{mn}$ will be used to raise and lower indices and to define a covariant derivative $D_m$.

\subsection{Proposing an effective action}
We start from the bosonic part of the Einstein-Maxwell action coupled to Abelian vector fields:
\begin{eqnarray}
S &=& \int \rmd^4 x \sqrt{g}\left(\ft12 R -g^{\mu \nu}G_{\alpha\bar
\beta}\partial_\mu z^\alpha \partial_\nu \bar z^{\bar \beta} + \ft14(\Im
\mathcal{N}_{IJ} )F_{\mu\nu}^I F^{\mu\nu J} -\ft1{8\sqrt{g}}(\Re \mathcal
N_{IJ})\veps^{\mu\nu\rho\sigma} F_{\mu\nu}^I
F_{\rho\sigma}^J\right)\nonumber\\
 &=& S_{\rm Einstein}+S^{(0)}+S^{(1)}\,.\label{S_one}
\end{eqnarray}
The independent variables in this action are the scalars $z^\alpha$,
gauge vectors $A^I_\mu$, appearing through their fields strengths
$F^I_{\mu\nu} = \partial_\mu A^I_\nu - \partial_\nu A^I_\mu$\,, and the
metric $g_{\mu\nu}$. We have split the action in a gravity term $S_{\rm
Einstein}$\,, a scalar part $S^{(0)}$ and a vector part $S^{(1)}$. When
seen in the context of a concrete $N=2$ supergravity theory, the scalar
metric $G_{\alpha\bar\beta}$ will be hermitian and linked to the K{\"a}hler
potential of the special geometry. The vector couplings are given by $\Im
\mathcal N_{IJ}$ and $\Re \mathcal N_{IJ}$. The matrix $\mathcal{N}$ is
determined by the special geometry and a choice of symplectic basis. The
matrix $\Im \mathcal N$ will be negative-definite. For our purposes,
however, the action may have both an arbitrary scalar metric
$G_{\alpha\bar\beta}$ and arbitrary scalar-dependent vector couplings
$\Im \mathcal N_{IJ}$ and $\Re \mathcal N_{IJ}$.

As explained in secetion \ref{sect-intro}, we would like to find an effective
action starting from expressions for the Einstein equations. The Einstein
tensor is defined as
\begin{equation}
 G_{\mu\nu} = 2(\sqrt{g})^{-1}\frac{\delta S_{\rm Einstein}}{\delta g^{\mu\nu}} = R_{\mu\nu}-\ft12g_{\mu\nu}R\,.
\end{equation}
The Einstein equations are given by varying the action
(\ref{S_one}) with respect to the metric and yield:
\begin{eqnarray}
G_{\mu\nu} = T_{\mu\nu}\,,\label{Einstein_eqs-formal}
\end{eqnarray}
where $T^{\mu\nu}$ is the energy-momentum tensor. We have
\begin{eqnarray}
  &&T_{\mu\nu} = T_{\mu\nu}^{(0)} + T_{\mu\nu}^{(1)}\,,\nonumber\\
  &&T^{(0)}_{\mu\nu} = -2(\sqrt{g})^{-1}\frac{\delta S^{(0)}}{\delta g^{\mu\nu}},\quad T^{(1)}_{\mu\nu} = -2(\sqrt{g})^{-1}\frac{\delta S^{(1)}}
  {\delta g^{\mu\nu}}\,.
\end{eqnarray}
 Remember that our solution will be stationary and hence time derivatives vanish. Let us
 look first at the contribution from the scalar part of the action. We have
\begin{equation}
  T_{\mu\nu}^{(0)} = G_{\alpha\bar\beta}(-g_{\mu\nu}g^{\rho\sigma}\partial_\rho z^{\alpha}\partial_\sigma\bar z^{\bar\beta} + 2\partial_\mu
  z^\alpha\partial_\nu\bar z^{\bar\beta})\,.\label{EM_scalars}
\end{equation}
For the vectors, we note that the metric only appears in the term with
$\Im \mathcal{N}$ in the supergravity action (\ref{S_one}). This leads to
\begin{equation}
T_{\mu\nu}^{(1)} = \Im \mathcal{N}_{IJ}\left(\ft14g_{\mu\nu}F_{\rho\sigma}^IF^{J\rho\sigma} - F_{\mu\rho}^IF_{\nu\sigma}^Jg^{\rho\sigma}\right) \label{T_munu}\,.
\end{equation}
When writing down the energy-momentum tensor, it proves useful to
introduce the magnetic vectors
\begin{equation}
  F^I_m=\ft12\gamma _{mn}(\sqrt{\gamma })^{-1}\veps ^{npq}F_{pq}^I.
 \label{magnvectors}
\end{equation}

After some calculation, we see that the Einstein equations reduce to the following two equations:
\begin{eqnarray}
 && -\ft12R_{mn}(\gamma )+\partial _mU\,\partial _n U + \partial _{(m}z^\alpha\partial _{n)}\bar
  z^{\bar\beta}-\rme^{2U}V_{mn}=0\,,\label{Einsteinconstraint}\\
  && D_m\partial ^m U -\rme^{2U}\gamma ^{mn}V_{mn}=0\,.\label{EinsteinFE}
\end{eqnarray}
The tensor $V_{mn}$ is quadratic in the field strengths and given by the expression:
\begin{equation}
 V_{mn}=-\ft12\Im\mathcal{N}_{IJ}\left(\rme^{-4U}F^I_{0m}F^J_{0n}+F_m^IF_n^J\right)\,.\label{gammaVBH}
\end{equation}

All this leads us to proposing the following effective action:
\begin{equation}
  S_{\rm eff}=\int \rmd t\int \rmd^3 x \sqrt{\gamma }\gamma ^{mn}\left[
  -\partial_mU\partial _n U -G_{\alpha\bar \beta}\partial_m z^\alpha \partial_n \bar z^{\bar \beta}-\rme^{2U}V_{mn}\right]
  \,.
 \label{EffectiveSgamma}
\end{equation}
The field equation of this action for $U$ is (\ref{EinsteinFE}) if we
keep $V_{mn}$ fixed during the variation. Of course, as $U$ does not
appear in the scalar part of the action, the latter is not determined by
this requirement. We will prove that the field equations for the scalars
can also be derived from this effective action for a specific
parametrization of $V_{mn}$. This will be clarified shortly. Only then
will it be clear how to use this effective action. We want to emphasize
that $\gamma^{mn}$ should not be seen as a dynamical variable in this
action. Instead of its field equations as following from this effective
action, we should take the Einstein equations from the original action
(\ref{Einsteinconstraint}) as constraints complementing the action
(\ref{EffectiveSgamma}). This will be explained below.

\subsection{Reproducing the field equations from the effective action}
In the effective action, expression (\ref{gammaVBH}) is phrased in
components of the field strengths $F_{\mu \nu }^I$. However, we can write
it in terms of the symplectic vectors of field strengths and field
equations. To do so, we introduce the following variables:
\begin{eqnarray}
  G_{I\mu \nu }&=&\veps_{\mu\nu\rho\sigma} \frac{\partial S}{\partial F_{\rho\sigma}^I}\nonumber\\
&=& (\Re{\cal N}_{IJ})F^J_{\mu \nu }+\ft12(\Im{\cal N}_{IJ})g_{\mu \mu
  '}g_{\nu \nu '}(\sqrt{g})^{-1}\varepsilon ^{\mu' \nu' \rho \sigma }F^J_{\rho \sigma}\,.
 \label{defGF}
\end{eqnarray}
Under electric/magnetic duality, $F^I$ and $G_I$ transform as a
symplectic vector. If we denote the three-dimensional duals as a
generalization of (\ref{magnvectors}):
\begin{equation}
  F^I_m=\ft12\gamma _{mn}(\sqrt{\gamma })^{-1}\varepsilon ^{npq}F_{pq}^I\,,\qquad
  G_{Im}=\ft12\gamma _{mn}(\sqrt{\gamma })^{-1}\varepsilon ^{npq}G_{Ipq}\,,
 \label{3ddualF}
\end{equation}
we obtain from (\ref{defGF}) and  $\varepsilon^{0npq}=-\varepsilon
^{npq}$ (as we use $\varepsilon _{0123}=-\varepsilon ^{0123}=1$):
\begin{equation}
  \begin{pmatrix}F_{0m}^I\cr G_{I0m}\end{pmatrix}=\rme^{2U}\Omega{\cal M}\begin{pmatrix}F^J_m\cr
  G_{Jm}\end{pmatrix}.
 \label{F0mFpq}
\end{equation}
The matrices in the expression are the symplectic metric $\Omega$ and the matrix $\mathcal M$ defined by:
\begin{equation}
  \Omega = \begin{pmatrix}0 &\unity\\-\unity &0 \end{pmatrix}\,,\quad {\cal M}=\begin{pmatrix}-I-RI^{-1}R&RI^{-1}\cr
  I^{-1}R&-I^{-1}\end{pmatrix}\,, \qquad R=\Re{\cal N}\,,\quad I=\Im{\cal
  N}\,.
 \label{OmMOm}
\end{equation}
These matrices contain indices $I$ and $J$ at appropriate positions
automatically for (\ref{F0mFpq}) to make sense. This leads to
\begin{eqnarray}
  V_{mn}&=& -\frac12\rme^{-2U}(F_m^IG_{I0n}-F_{0m}^IG_{In})\nonumber\\
  &=&\frac{1}{2}\begin{pmatrix}F^I_m & G_{Im}\end{pmatrix}
  {\cal M}\begin{pmatrix}F^J_n\cr G_{Jn}\end{pmatrix}\,.
 \label{Vmnsympl}
\end{eqnarray}

$U$ does not appear in this expression for $V_{mn}$. This implies that if
we consider $V_{mn}$ as a function of $F^I_m$, $G_{Im}$ and the scalars
implicitly present in (\ref{OmMOm}) and insert it as such in the
effective action (\ref{EffectiveSgamma}), then it is still valid that
this action generates the right field equation for $U$. We now check that
in this way it also generates the same scalar field equations as those
obtained from $S^{(0)}$ and $S^{(1)}$ in the original action
(\ref{S_one}), where the vector fields $A_\mu ^I$ were the other
independent variables. Hence the field equations that should be
reproduced are
\begin{equation}
 0= \partial _\mu (\sqrt{g}g^{\mu \nu }G_{\alpha\bar \beta}\partial_\nu\bar z^{\bar \beta})+\ft14\sqrt{g}\partial _\alpha(\Im {\cal
N}_{IJ})F_{\mu\nu}^I F^{\mu\nu J} -\ft 18 \partial _\alpha(\Re {\cal N}_{IJ})
\varepsilon^{\mu\nu\rho\sigma}F_{\mu\nu}^I F_{\rho\sigma}^J\,,
 \label{scalarfieldeqnorig}
\end{equation}
where we denoted $\partial_\alpha = \frac{\partial}{\partial
z^{\alpha}}$. Specifying metric (\ref{stationaryMetric}) and the
expressions for the field strengths in terms of $F^I_m$ and $G_{Im}$,
this becomes
\begin{equation}
 0= \partial _m(\sqrt{\gamma }\gamma ^{mn}G_{\alpha\bar \beta}\partial_n\bar z^{\bar \beta})-\frac{1}{2}\rme^{2U}\sqrt{\gamma
  }\gamma ^{mn}\begin{pmatrix}F^I_m&G_{mI}\end{pmatrix}
\partial_\alpha {\cal M} \begin{pmatrix}F^J_n\cr
G_{nJ}\end{pmatrix}\,,
 \label{scalarfieldeqn2}
\end{equation}
where the indices $I$ and $J$ appear again in appropriate positions on
the submatrices of $\partial_\alpha {\cal M}$. The latter are indeed the
field equations for $z^\alpha$, obtained from the effective action
\begin{equation}
   S_{\rm eff}=\int \rmd t\int \rmd^3 x \sqrt{\gamma }\gamma ^{mn}\left[
  -\partial_mU\,\partial _n U -G_{\alpha\bar \beta}\partial_m z^\alpha \partial_n \bar z^{\bar \beta} -\frac{1}{2}\rme^{2U}\begin{pmatrix}F^I_m & G_{Im}\end{pmatrix}{\cal M}\begin{pmatrix}F^J_n\cr G_{Jn}\end{pmatrix}\right]
  \,.
 \label{EffectiveSgammaUz}
\end{equation}
This action should be considered as an effective action for varying with
respect to the variables $U$, $z^\alpha$ and $A_m^I,A_{Im}$, while
$\gamma ^{mn}$ should be considered as background. We saw already that
the field equations of the original action for $\gamma ^{mn}$ lead to
constraint (\ref{Einsteinconstraint}). We will now check what the field
equations of the vector sector impose.

The Bianchi identities and the vector field equations from the
four-dimensional action (\ref{S_one}) with independent vectors $A_\mu ^I$
are equivalent to
\begin{equation}
  \varepsilon ^{\mu \nu \rho \sigma }\partial _\nu \begin{pmatrix}F^I_{\rho \sigma }\cr G_{I\rho \sigma
  }\end{pmatrix}=0\,.
 \label{FEBianchi}
\end{equation}
We can reproduce these equations from the effective action
(\ref{EffectiveSgammaUz}), if we now take three-dimensional gauge vectors
$A_m^I$, $A_{Im}$ as the fundamental vector variables. We consider then
$F_{mn}^I$ and $G_{Imn}$ as the field strengths for these variables:
\begin{equation}
 F_{mn}^I = \partial_mA^I_n - \partial_nA^I_m,\qquad G_{Imn} = \partial_mA_{In} - \partial_nA_{Im}\,.
\end{equation}
The Bianchi identities for the vectors $A_m^I,A_{Im}$ are given by
\begin{equation}
  \varepsilon ^{mnp}\partial _m \begin{pmatrix}F^I_{np}\cr G_{Inp
  }\end{pmatrix}=0\,.
 \label{Bianchi2}
\end{equation}
 The components $F_{0m}^I, G_{I0m}$ are no longer independent variables. Indeed, expression (\ref{F0mFpq}) should now be seen as their defining relation, showing how we can construct them from the three-dimensional vectors. Then the field equations for the vectors $A_m^I$, $A_{Im}$ obtained from varying the effective action (\ref{EffectiveSgammaUz}) can be written as
\begin{equation}
  \varepsilon ^{mnp}\partial _n \begin{pmatrix}F^I_{0p}\cr G_{I0p}\end{pmatrix}=0\,.
 \label{Fieldeq_vect}
\end{equation}
 We see that, for stationary solutions, the vector field equations and Bianchi identities (\ref{FEBianchi}) are correctly reproduced by (\ref{Bianchi2}) and (\ref{Fieldeq_vect}).

The effective action we propose (\ref{EffectiveSgammaUz}) indeed
reproduces all the field equations, with the following independent
fields: scalars $U$ and $z^\alpha$, and three-dimensional vectors
$A_{Im}$, $A^I_m$. To incorporate the full set of Einstein equations,
constraint (\ref{Einsteinconstraint}) still has to be considered. Note
that we arrive at an action which is no longer covariant in
four-dimensional spacetime, due to the vector term. We do however have an
action that displays invariance under electric/magnetic duality. We come
back to this in section \ref{sect-comparing_actions}.

\subsection{The black hole potential\label{ss_BHpotential}}
We now wish to establish the link with the black hole potential known in the literature. To do this, we try to find a solution to the equations  (\ref{FEBianchi}). In terms of our preferred variables, this gives:
\begin{equation}
 \partial _m\sqrt{\gamma
}\gamma ^{mn} \begin{pmatrix}F^I_{n}\cr G_{In}\end{pmatrix}=0, \qquad
\partial _{[m}\rme^{2U}{\cal M}\Omega\begin{pmatrix}F^I_{n]}\cr
G_{In]}\end{pmatrix}=0.
 \label{FEBianchi0m}
\end{equation}

One way of solving these equations is to put
\begin{eqnarray}
  &&F^I_m=\partial _mH^I\,,\qquad G_{Im}=\partial _mH_I\,, \nonumber\\
&&\partial _m  \sqrt{\gamma }\gamma ^{mn}\partial _n H^I=\partial _m
\sqrt{\gamma }\gamma ^{mn}\partial _n
  H_I=0\,.
 \label{harmonicH}
\end{eqnarray}
Remark that this brings us in the realm of supersymmetric solutions.
Indeed, for instance in \cite{Denef:2000nb} it was shown that for BPS
solutions the three-dimensional vectors are related to harmonic functions
on $\mathbb{R}^3$ precisely in the manner (\ref{harmonicH}).

We remain then with Bianchi identities of the form
\begin{equation}
  \left( \partial _{[m}\rme^{2U}{\cal M}\right)\Omega\partial _{n]} \begin{pmatrix}H^I\cr
H_I\end{pmatrix}=0\,,
 \label{RestBianchi}
\end{equation}
which can be solved by assuming that all functions ($U$, the scalars and
the harmonic $H^I$ and $H_I$) depend only on one coordinate such that the
$\partial _m$ and $\partial _n$ for $m\neq n$ in the above equation
cannot both be non-vanishing. We denote this one coordinate as $\tau $.
Thus, we have: $U(\tau )$, $z(\tau )$, $H^I(\tau )$ and $H_I(\tau )$.

A convenient metric is e.g. \cite{Rasheed:1997ns,Gibbons:1996af}
\begin{equation}
\gamma _{mn}\rmd x^m\rmd x^n = \frac{c^4}{\sinh^4 c\tau}\rmd\tau^2 +
\frac{c^2}{\sinh^2 c\tau}(\rmd\theta ^2 + \sin^2\theta \rmd\varphi ^2)\,.
 \label{sphericalcoord}
\end{equation}
Details on this parametrization are given in an appendix of
\cite{Kallosh:2006bt}. This parametrization has the property
$\sqrt{\gamma }\gamma ^{\tau \tau }=\sin\theta $, which will be useful.

Harmonic means now just $\ddot H=0$, where a dot is a derivative w.r.t.
$\tau $. So we can take
\begin{equation}
H=  \begin{pmatrix}H^I\cr H_I\end{pmatrix}= \Gamma \tau +h,\qquad
h=\begin{pmatrix}h^I\cr h_I\end{pmatrix}, \qquad \Gamma
=\begin{pmatrix}p^I\cr q_I\end{pmatrix}.
 \label{Hpqh}
\end{equation}
We have here introduced the magnetic and electric charges in the
symplectic vector $\Gamma $.

In this parametrization, the Lagrangian gets the form (up to a normalization):
\begin{equation}
  {\mathcal L}_{\rm eff}=\dot U^2 +G_{\alpha\bar\beta}\dot z^\alpha\dot{\bar z}^{\bar\beta}+\rme^{2U}V_{\rm BH}\,,
 \label{SeffU}
\end{equation}
where the black hole potential is
\begin{equation}
  V_{\rm BH}=V_{\tau \tau }=\ft12\Gamma ^T{\cal M}\Gamma. \label{VBH}
\end{equation}
The above is also the form of the reduced supergravity action, obtained
by integrating out angular dependence, known in the literature (see e.g.
\cite{Ferrara:1997tw,Kallosh:2006bt}).

The one-dimensional effective Lagrangian (\ref{SeffU}) does not reproduce
all of the field equations. Indeed, from the Einstein equations
(\ref{Einsteinconstraint}), we have to impose the following constraint:
\begin{equation}
c^2-\dot U^2-G_{\alpha\bar\beta}\dot z^\alpha\dot{\bar z}^{\bar\beta}+\rme^{2U}V_{\rm BH}=0\,.
 \label{IndepEinstein}
\end{equation}

\section{Non-static metric\label{sect-nonstatic}}
Having established the key ideas in the last section, we would like to
extend this now to the case where we have a more general, non-static
ansatz for the metric. In \cite{Tod:1983pm,Tod:1995jf} it was shown that
time independent (BPS) configurations require a stationary metric that
can be written in the form:
\begin{equation}
  \rmd s^2 = -\rme^{2U}(\rmd t + \omega_m \rmd x^m)^2 + \rme^{-2U} \gamma_{mn}\rmd x^m \rmd x^n\,.\label{g_multi}
\end{equation}
In fact, \cite{Tod:1983pm,Tod:1995jf} dictates $\gamma_{mn} =
\delta_{mn}$. We will extend this slightly however, and allow
$\gamma_{mn}$ to be an arbitrary metric on the three-dimensional space.
We will proceed following the same steps as for a static metric. The main
objective is to generalize the concept of the black hole potential, which
was the tensor $V_{mn}$ (\ref{Vmnsympl}) written down in a specific
spherical coordinate system, and find the generalization of the extra
constraint (\ref{Einsteinconstraint}). We will keep using $\gamma_{mn}$
to raise an lower indices and to define a covariant derivative $D_m$.
Again, we will  make use of the dual field strengths (\ref{defGF}).
Remember, however, that this definition will lead to different
expressions than in the static case, because of the explicit metric
dependence in (\ref{defGF}). Also, we define the three-dimensional duals
as in (\ref{3ddualF}).

In subsection \ref{SS_fieldeqs} we give the field equations that have to
be reproduced by the effective action we will construct. We calculate the
Einstein equations and the field equations for the scalars and the
vectors. We will later propose an effective action, in three-dimensional
flat space. The function $U$ appearing in the metric will show up in this
action as a scalar field; the functions $\omega_m$ behave as
three-dimensional gauge vectors. Anticipating this, we will introduce the
following field strength for $\omega_m$:
\begin{equation}
 \Omega_{mn}\equiv \partial_{m}\omega_n - \partial_n\omega_m\,.
\end{equation}
We will also find several constraints that have to be imposed on the
fields in the effective action. Later, in subsection
\ref{SS_newvariables}, we cast the effective action in a convenient form
using variables $G_{I\mu\nu}$. We find in particular how the black hole
potential can be generalized. The independent fields in the effective
action will be the metric components $U$ and $\omega$, the scalars
$z^\alpha$ and three-dimensional gauge vectors $A_m^I$, $A_{Im}$. It is
then easy to show that the effective action correctly reproduces the
field equations for the scalar fields and the vector field equations of
the original supergravity action (\ref{S_one}).

\subsection{Field equations}\label{SS_fieldeqs}
We would like to construct an effective action, which correctly describes
the dynamics of the $N=2$ supergravity theory, encoded in  the action
(\ref{S_one}). Concretely, this means that we want to reproduce the
scalar and vector field equations, and (part of) the Einstein equations.
The metric ansatz we use now is the non-static one, given by
(\ref{g_multi}). Again, we restrict ourselves to stationary solutions.

The scalar field equations for the supergravity action (\ref{S_one}) are
given by
\begin{equation}
 0= \partial _\mu (\sqrt{g}g^{\mu \nu }G_{\alpha\bar \beta}\partial_\nu\bar z^{\bar \beta})+\ft14\sqrt{g}\partial _\alpha(\Im {\cal
N}_{IJ})F_{\mu\nu}^I F^{\mu\nu J} -\ft 18 \partial _\alpha(\Re {\cal
N}_{IJ}) \varepsilon^{\mu\nu\rho\sigma}F_{\mu\nu}^I F_{\rho\sigma}^J\,.
 \label{scalarfieldeqnorig2}
\end{equation}
Expressing everything in terms of vectors $F_{m}^I$ and $G_{Im}$ using
(\ref{3ddualF}), this is equivalent to
\begin{eqnarray}
  0  &=&D^m (G_{\alpha\bar \beta}\partial_m\bar z^{\bar \beta})\nonumber\\
&&-\frac{1}{2}\frac{\rme^{2U}}{\rme^{4U}\omega^2-1} \Big{[}(\rme^{4U}\omega^m\omega^n-\gamma^{mn})\begin{pmatrix}F_m &
  G_{m}\end{pmatrix} \partial_\alpha{\cal M}\begin{pmatrix}F_n\cr G_{n}\end{pmatrix}\Big{]}\,.\label{scalarfieldeqs}
\end{eqnarray}
The vector field equations and Bianchi identities on the other hand, can
be written as
\begin{equation}
  \varepsilon ^{\mu \nu \rho \sigma }\partial _\nu \begin{pmatrix}F^I_{\rho \sigma }\cr G_{I\rho \sigma
  }\end{pmatrix}=0\,.
 \label{bianchis}
\end{equation}

To find the Einstein equations, we start from the components of the Einstein tensor:
\begin{eqnarray}
G_{00} &=& \rme^{4U}(\ft12 R(\gamma)+2{ D}_n\partial^n U - \partial_n U \partial^n U + \ft38 e^{4U}\Omega_{mn}\Omega^{mn})\,,\nonumber\\
G_{m0} &=& \omega_m G_{00} - \ft12 {D}^p \left(\rme^{4U}\Omega_{mp}\right)\,,\nonumber\\
G_{mn} &=& \omega_m\omega_n G_{00}- \omega_{(m}D^p \left(\rme^{4U}\Omega_{n)p)}\right)\label{Einstein_tensor}\\
&&+\left(\gamma_{mn}\gamma^{pq} -
2\delta_{(m}^p\delta_{n)}^{q}\right)\left(-\ft12 R_{pq}(\gamma) +
\partial_p U\partial_{q} U \right) - \ft18\left(\gamma_{mn}\gamma^{pq} -
4\delta_{(m}^p\delta_{n)}^{q}\right) \rme^{4U}\Omega_{pr}\Omega_{q}{}^{r}
\,.\nonumber
\end{eqnarray}
To calculate the components of the energy-momentum tensor, we again make
use of expressions (\ref{EM_scalars}) and (\ref{T_munu}), but now for the
non-static metric ansatz (\ref{g_multi}). Again, we make use of magnetic
vectors $F^I_m$ as defined in  (\ref{3ddualF}). We have a similar
definition for $V_{mn}$ as in (\ref{gammaVBH}) and in addition, we define
another symmetric tensor $A_{mn}$, quadratic in the field strengths:
\begin{eqnarray}
V_{mn} &=& -\ft12 (\Im\N_{IJ}) \left(\rme^{-4U}F^I_{0m}F^J_{0n}+  F^I_{m} F^{J}_n\right)\label{V_mn}\,,\\
A_{mn} &=& -\ft14 (\Im\N_{IJ}) \left(\omega_r\omega^r F^I_{0m}F^I_{0n}+\omega_m\omega_nF^I_{0r}F^{J}_0{}^r-2\omega_p\omega_{(m}F^I_{0}{}^pF^I_{0n)} \right.\nonumber\\
&& \left.- 2\veps_{(m|pq|}\omega_{n)}
F^I_{0}{}^pF^{Jq}-2\veps_{p(m|q}\omega^pF^I_{0|n)}{}F^{Jq}\right)\,.\label{A_mn}
\end{eqnarray}

Imposing that each of the Einstein equations should vanish, we can in fact show that they are equivalent to the following set of independent equations:
\begin{eqnarray}
0&=&2D_n\partial^n U - 2\rme^{2U}\gamma^{mn}(V_{mn}+A_{mn}) +\ft12\rme^{4U}\Omega_{mn}\Omega^{mn}\,, \label{multi-U_feqn}\\
0 &=&\ft12 D^r \left(\rme^{4U}\Omega_{mr}\right)
+\rme^{2U}(\Im \mathcal{N}_{IJ})\left(\omega_mF^I_{0p}F^{J}_0{}^p - \omega_p F_0^I{}^pF^J_{0m}-\veps_{mpq}F^I_{0}{}^pF^{Jq} \right)\,,\label{multi-om_feqn}\\
0 &=&\left(\gamma_{mn}\gamma^{pq} - 2\delta_{m}^p\delta_{n}^q\right)\left(-\ft12 R_{pq}(\gamma) +\partial_p U\partial_q U +\partial_p  z^{\alpha}\partial_q \bar z^{\bar\beta}-\rme^{2U}V_{pq}\right)\nonumber\\
&&+\left(\gamma_{mn}\gamma^{pq} - 4\delta_{m}^p\delta_{n}^q\right)(\rme^{2U
}A_{pq} -\ft18\rme^{4U}\Omega_{pr}\Omega_q{}^{r})\,.  \label{multi-constraint}
\end{eqnarray}
Now we will propose an effective action that correctly describes the above dynamics.

\subsection{The effective action\label{SS_newvariables}}
The effective action we propose will have the concise form
\begin{equation}
S_{\rm eff} = \int \rmd t\int \rmd^3 x \sqrt{\gamma}\left[- \partial_m
U\partial^m U - G_{\alpha\bar \beta}\partial_m z^\alpha \partial^m \bar
z^{\bar \beta} +\ft18\rme^{4U}\Omega_{mn}\Omega^{mn}\right] +\int \rmd
t\int \rmd^3 x\,  {\mathcal L}_{\rm eff}^{(1)}\,. \label{S_multi}
\end{equation}

Now we would like to cast the action (\ref{S_multi}) in a form similar to
that we derived for the static solution. In particular, we want to extend
the concept of a black hole potential we introduced there. To do this, we
determine what the vector part of the effective action, ${\mathcal
L}_{\rm eff}^{(1)}$,  should look like, to reproduce the scalar field
equations (\ref{scalarfieldeqs}) as well as the field equations and
Bianchi identities for the vectors (\ref{bianchis}). First, we will
choose as dynamical vector variables in the effective action
(\ref{S_multi}) vectors $A_m^I$, $A_{Im}$, as was done in the static
case, with
\begin{equation}
F_{mn}^I =   \partial_mA_n^I - \partial_nA_m^I\,,\qquad G_{Imn} =   \partial_mA_{In} - \partial_nA_{Im}\,.
\end{equation}
{}From (\ref{defGF}), we obtain that the vectors $F_{0m}^I$ are functions
of the independent variables we have in the effective action (namely $U$,
$\omega_m$, $z^\alpha$ and $A_m^I,A_{Im}$)\,:
\begin{equation}
  \begin{pmatrix}F_{0m}^I\cr G_{I0m}\end{pmatrix} =
  \frac{\rme^{2U}}{\rme^{4U}\omega^2-1}\Big{[}(\rme^{4U}\omega_m\omega_n
  -\gamma_{mn})\Omega {\cal M}
   +(\sqrt{\gamma})^{-1}\gamma_{mq}\rme^{2U}\veps^{npq}\omega_p\unity \Big{]}\begin{pmatrix}F^{J}_n\cr G_{Jn}\end{pmatrix}\,,\label{F_0m-G_0m1}
\end{equation}
where the indices $I$ and $J$ are suppressed on the matrices $(\Omega
{\cal M})$ and $\unity $. They appear in the different entries
automatically in the appropriate positions.

The expression (\ref{Vmnsympl}) for $V_{mn}$ in the static case, suggests
we take the following expression as the vector-dependent part in the
effective action (\ref{S_multi}):
\begin{eqnarray}
  (\sqrt{\gamma})^{-1}{\mathcal L}_{\rm eff}^{(1)} &=&\frac12(F^{Im}G_{I0m} -G_{I}^mF^{I}_{0m})\nonumber\\
  &=&  -\frac12\frac{\rme^{2U}}{\rme^{4U}\omega^2-1} \Big{[}(\rme^{4U}\omega^m\omega^n-\gamma^{mn})\begin{pmatrix}F^I_m &
  G_{Im}\end{pmatrix}  {\cal M}\begin{pmatrix}F^J_n\cr G_{Jn}\end{pmatrix}\nonumber\\
&&+2(\sqrt{\gamma})^{-1}\rme^{2U}\omega_mG_{In}F_p^I\veps^{mnp}\Big{]}\label{FG0_minus_GF0}\,.
\end{eqnarray}
Now we will calculate the field equations for the independent fields in the effective action (\ref{S_multi}), showing why the choice (\ref{FG0_minus_GF0}) is justified.

First, we can take a variation of the effective action w.r.t. the scalar
field $U$. Using (\ref{F_0m-G_0m1}), we find
\begin{eqnarray}
  \frac{ \delta {\mathcal L}_{\rm eff}^{(1)}}{\delta U} &=&
  \sqrt{\gamma}\rme^{2U}\Im\N_{IJ}\Big{(}F^I_{m} F^{Jm}+\rme^{-4U}F^I_{0m}F^J_{0}{}^m  +\Omega_{mn}F^I_{0}{}^mF^{J}_0{}^n- 2(\sqrt{\gamma})^{-1}\veps^{mnp}\omega_mF_{0n}^IF^{J}_p\Big{)}\nonumber\\
  &=& -2\sqrt{\gamma}\rme^{2U}\gamma^{mn}(V_{mn}+A_{mn})\,.
\end{eqnarray}
The functional derivative on the left-hand side has to be carried out
when keeping the fields $\omega_m$, $z^\alpha$ and $A_m^I,A_{Im}$ fixed.
When we now make a variation of the full effective action
(\ref{S_multi}), we see that the field equation for $U$ is the same as
the Einstein equation (\ref{multi-U_feqn}). Doing a likewise variation to
find the field equations for $\omega_m$, keeping $F_m^I,G_{Im}$ fixed
during variation, we reproduce expression (\ref{multi-om_feqn}). A
variation w.r.t. $\gamma_{mn}$ gives inconsistency with the Einstein
equations (\ref{multi-U_feqn})--(\ref{multi-constraint}).

We now see that the effective action also generates the same scalar field equations as those obtained from the original action (\ref{S_one}), where the vector fields $A_\mu ^I$ were the other independent variables. Indeed, if we take the vectors $A_m^I$, $A_{Im}$ fixed during variation, the scalar field equations from the effective action (\ref{S_multi}) are exactly given by (\ref{scalarfieldeqs}).

Also the vector field equations and Bianchi identities (\ref{bianchis})
are reproduced. The Bianchi identities for gauge vectors $A_m^I$,
$A_{Im}$ are given by
\begin{equation}
  0 = \veps^{mnp}\begin{pmatrix}\partial_m F_{Inp}\cr
  \partial_m G_{Inp}\end{pmatrix}\,.\label{bianchis2}
\end{equation}
Using (\ref{F_0m-G_0m1}) as a definition for $F_{0m}^I$ and $G_{I0m}$,
the field equations that follow from the effective action (\ref{S_multi})
are given by
\begin{equation}
  0 =\veps^{mpq}\partial_p\begin{pmatrix}F_{0q}^I\cr G_{I0q}\end{pmatrix}\,.\label{multi_fieldeqs}
\end{equation}
We see that for stationary solution ($\partial_0F_{mn} = 0$), the field
equations and Bianchi identities of the four-dimensional action
(\ref{S_one}) are correctly reproduced by (\ref{bianchis2}) and
(\ref{multi_fieldeqs}).

Finally we note that the Einstein equations (\ref{multi-constraint})
cannot be reproduced from the effective action we propose. The conclusion
is that they have to be supplied as additional constraints on the fields
appearing in the effective action (\ref{S_multi}). They read
\begin{eqnarray}
0 &=&\left(\gamma_{mn}\gamma^{pq} - 2\delta_{m}^p\delta_{n}^q\right)\left(-\ft12 R_{pq}(\gamma) + \partial_p U\partial_q U +\partial_p  z^{\alpha}\partial_q \bar z^{\bar\beta}-\rme^{2U}V_{pq}\right)\nonumber\\
&&+\left(\gamma_{mn}\gamma^{rs} - 4\delta_{m}^r\delta_{n}^s\right)(\rme^{2U
}A_{pq} -\ft18\rme^{4U}\Omega_{pr}\Omega_q{}^{r})\,,  \label{multi-constraint2}
\end{eqnarray}
where $V_{pq}$ and $A_{pq}$ where given in (\ref{V_mn}) and (\ref{A_mn}).
The $F_{0m}$ appearing in these terms has to be understood as a function
of $F_{m}^I$ and $G_{Im}$ through relation (\ref{F_0m-G_0m1}).

We have seen that the effective action (\ref{S_multi}) correctly
reproduces the field equations, for our non-static metric choice.  It is
important to note that the three-dimensional metric $\gamma_{mn}$ has to
be considered as a background field in the effective action. Some of the
Einstein equations cannot be reproduced -- they have to be taken along as
extra constraints. Again, the idea of four-dimensional gauge vectors has
to be abandoned. The variables to be considered now are three-dimensional
vectors $A^I_m$, $A_{Im}$. This implies that the effective action is no
longer covariant. It does however remain invariant under
electric/magnetic duality rotations. More on this in the following
section. In the light of the discussion for non-static metrics, we also
have a generalization of the concept of the black hole potential. It is
given by the vector Lagrangian (\ref{FG0_minus_GF0}).

\section{Electric/magnetic duality\label{sect-comparing_actions}}
We would like to show that the supergravity action (\ref{S_one}) we
started from  is not equivalent to the effective action (\ref{S_multi})
we constructed. In fact, this can be seen from electric/magnetic duality.
It is known that the Lagrangian (\ref{S_one}) is not invariant under
electric/magnetic duality. However, the effective action (\ref{S_multi})
is invariant.

In a first section, we show this in detail, while in a second subsection
we establish a link with the work of Denef \cite{Denef:2000nb} and the
entropy function as introduced in \cite{Sen:2005wa,Cardoso:2006xz}. In
particular, the effective action (\ref{SeffU}) we obtained for
spherically symmetric metrics, gives exactly that entropy function.

\subsection{Inequivalence of the `original' and effective actions}
\paragraph{Non-static metric.}
On the one hand, we have the supergravity action (\ref{S_one}):
\begin{equation}
S = \int \rmd^4 x \sqrt{g}\left(\ft12 R -g^{\mu \nu}G_{\alpha\bar \beta}\partial_\mu z^\alpha \partial_\nu \bar z^{\bar \beta} + \ft14(\Im \mathcal{N}_{IJ} )F_{\mu\nu}^I F^{\mu\nu J} -\ft1{8\sqrt{g}}(\Re \mathcal N_{IJ})\veps^{\mu\nu\rho\sigma} F_{\mu\nu}^I F_{\rho\sigma}^J\right)\,.\label{S_onebis}
\end{equation}
The Ricci scalar can be readily calculated from (\ref{Einstein_tensor}) and reads:
\begin{eqnarray}
  R &=& - g^{\mu\nu}G_{\mu\nu}\nonumber\\
  &=& \rme^{2U}(R(\gamma) -2\partial_mU\partial^mU +\ft14\rme^{4U}{\Omega}_{mn}{\Omega}^{mn} -2\partial_m\partial^m U)\,.
\end{eqnarray}
Again, we used the three-dimensional metric $\gamma _{mn}$ to raise and
lower indices. We also use the expressions for $G_{I\mu\nu}$
(\ref{defGF}) and for the vectors $F_m^I$, $G_{Im}$ (\ref{3ddualF}) to
rewrite the vector part of the Lagrangian:
 \begin{eqnarray}
  {\mathcal L}^{(1)} &\equiv& \ft14\sqrt{g}(\Im \mathcal{N}_{IJ} )F_{\mu\nu}^I F^{\mu\nu J} -\ft1{8}(\Re \mathcal N_{IJ})\veps^{\mu\nu\rho\sigma} F_{\mu\nu}^I F_{\rho\sigma}^J\nonumber\\
  &=&-\frac1{8}\veps^{\mu\nu\rho\sigma}F_{\mu\nu}G_{\rho\sigma}\nonumber\\
  &=& \frac12\sqrt{\gamma}(F^{Im}G_{I0m} + G_I^mF^I_{0m})\,.\label{L1}
\end{eqnarray}
Specifying the metric and restricting ourselves again to stationary
solutions, the supergravity action is, up to a total derivative, given by
\begin{equation}
S = \int \rmd^4 x
\sqrt{\gamma}\left[\ft12R(\gamma)-\partial_mU\partial^mU
+\ft18\rme^{4U}{\Omega}_{mn}{\Omega}^{mn} - G_{\alpha\bar
\beta}\partial_m z^\alpha \partial^m \bar z^{\bar \beta} \right] + \int
\rmd^4 x\,{\mathcal L}^{(1)}\,.\label{S_red}
\end{equation}
This action is not the same as the effective action we obtained earlier.
First, it has to be understood in a different way -- the reduced action
(\ref{S_red}) has as independent variables $U$, $\omega$, the scalars
$z^\alpha$ and four-dimensional gauge vectors $A_\mu^I$. In this way it
reproduces the field equations of the scalars and the vectors, and the
Einstein equations (\ref{multi-U_feqn}) and (\ref{multi-om_feqn}). Again,
(\ref{multi-constraint}) has to be taken along as an extra constraint.
But more importantly, it is clear that the vector Lagrangian is not
invariant under electric/magnetic duality.

This is clearly different from the effective action (\ref{S_multi}) we obtained. There we had instead of ${\mathcal L}^{(1)}$ the following vector part in the Lagrangian:
\begin{eqnarray}
  {\mathcal L}_{\rm eff}^{(1)} &=& \frac12\sqrt{\gamma}(F^{Im}G_{I0m}-G_{I}^mF_{0m}^I)\,,\label{L_effGF}
\end{eqnarray}
the independent vector variables being three-dimensional gauge vectors
$A^I_m$ and $A_{Im}$. This expression clearly is invariant under
electric/magnetic duality rotations.
\paragraph{Static metric.} When we use the metric ansatz (\ref{stationaryMetric}), we can carry through the same discussion, with $\omega_m = 0$.
\subsection{Discussion and comparison to the literature}
The expression (\ref{L_effGF}) is truly invariant under electric/magnetic
duality rotations. This shows that our method of constructing an
effective action leads us from a form (\ref{L1}) which makes general
covariance manifest, but lacking electric/magnetic invariance, to a form
(\ref{L_effGF}) which incorporates the latter but misses out on
four-dimensional covariance. In fact, a manifestly covariant
four-dimensional action invariant under electric/magnetic duality does
not exist. In \cite{Denef:2000nb}, Denef comments on this. Because he
works in the framework of type IIB compactification, the problem is
equivalent to the non-existence of a straightforward generally covariant
action for the self-dual four-form potential. Based on the work of
\cite{Henneaux:1988gg,Bekaert:1998yp}, he constructs a four-dimensional
action which exhibits invariance under electric/magnetic duality
rotations. Our effective action is in fact the same as his, up to a
normalization of the vectors. However, in his work constraints
(\ref{multi-constraint2}) were not taken into account.

We establish the connection with the entropy function in the literature \cite{Sen:2005wa,Cardoso:2006xz}. To do so, we write the Lagrangian  of the original action  (\ref{S_red}) in a suggestive form:
\begin{equation}
  {\mathcal L}(F_{0m}^I,F_m^I,X^a,z^\alpha)\,. \label{L}
\end{equation}
Here $X^a$ is the collection of dynamical fields in the metric: $U$ for
the static metric (\ref{stationaryMetric}), while in the case of the
non-static metric (\ref{g_multi}) it is given by $U$ and $\omega_m$. Note
that we have to consider the three-dimensional metric $\gamma$ as
background.

{}From the definition of $G_{I\mu\nu}$, or equivalently, from (\ref{L1}),
we see that
\begin{eqnarray}
 G_{I0m} = (\sqrt{\gamma})^{-1}\gamma_{mn}\frac{\partial {\mathcal L}}{\partial F^I_n}, \qquad G_{Im} = (\sqrt{\gamma})^{-1}\gamma_{mn}\frac{\partial {\mathcal L}}{\partial F^I_{0n}}\label{Leg_transform}\,.
\end{eqnarray}
Then (\ref{S_red}) and (\ref{L_effGF}) show that the effective Lagrangian
we obtained can be written as
\begin{equation}
  {\mathcal L}_{\rm eff}(F_m^I,G_m^I, X^a,z^\alpha) = {\mathcal L}(F_m^I,F_{0m}^I,  X^a,z^\alpha) - \sqrt{\gamma}F^I_{0m}G_I^m
-  \ft12\sqrt{\gamma}R(\gamma)
  \,.\label{L_eff-F}
\end{equation}
The last term can be considered as a constant because $\gamma $ is not a
dynamical field. Apart from this term,  (\ref{L_eff-F}) clearly takes the
form of a Legendre transform, where we switch from a description in
$F_m^I,F_{0m}^I$ (with gauge vectors $A_\mu^I$) to $F_m^I,G_m^I$ (with
gauge vectors $A^I_m,A_{Im}$). By general principles of Legendre
transformations this proves again that the field equations for $X^a$ and
$z^\alpha $ of ${\cal L}_{\rm eff}$ and of ${\cal L}$ are identical.

We can now carry this through in the spherically symmetric case. When we
use the parametrization of the metric as in  (\ref{sphericalcoord}), and
use the solution discussed in section \ref{ss_BHpotential}, the only
nonzero components of the field strengths $F^I_m$ and $G_{Im}$ are
\begin{equation}
 F_\tau^I = p^I,\qquad G_{I\tau} = q_I\,,
\end{equation}
After integrating out the angular dependence, (\ref{L_eff-F}) becomes the entropy function in the literature,
derived from the 2-derivative Lagrangian (\ref{L1}).

\section{Concluding remarks\label{sect-conclusion}}
In this paper, we started from the nicely covariant action (\ref{S_one}),
describing Einstein-Maxwell theory for a number of Abelian gauge vectors.
We investigated the field equations for stationary solutions. In
particular, we showed that the field equations can be reproduced from an
action that lacks in covariance, but makes electric/magnetic duality
manifest. The key idea was to shift from four-dimensional gauge vectors
$A_\mu^I$ to three-dimensional ones: $A_{m}^I,A_{Im}$. This effective
action is given by
\begin{equation}
S_{\rm eff} = \int \rmd t\int \rmd^3 x \sqrt{\gamma}\left[- \partial_m
U\partial^m U - G_{\alpha\bar \beta}\partial_m z^\alpha \partial^m \bar
z^{\bar \beta} +\ft18\rme^{4U}\Omega_{mn}\Omega^{mn}\right]  + \int \rmd
t\int \rmd^3 x\,{\mathcal L}_{\rm eff}^{(1)}\,,
\end{equation}
with
\begin{eqnarray}
 {\mathcal L}_{\rm eff}^{(1)} &=&\frac12\sqrt{\gamma}(F^{Im}G_{I0m} -G_{I}^mF^{I}_{0m})\\
  &=&  -\frac12\frac{\rme^{2U}}{\rme^{4U}\omega^2-1} \Big{[}\sqrt\gamma(\rme^{4U}\omega^m\omega^n-\gamma^{mn})\begin{pmatrix}F^I_m &
  G_{Im}\end{pmatrix}  {\cal M}\begin{pmatrix}F^J_n\cr G_{Jn}\end{pmatrix}+2\rme^{2U}\omega_mG_{In}F_p^I\veps^{mnp}\Big{]}\nonumber\,.
\end{eqnarray}
$F^I_m$ and  $G_{Im}$ are given by (\ref{3ddualF}). This action is
actually a Legendre transformation of the spacetime covariant
supergravity action (\ref{S_one}) on the level of the (three-dimensional)
gauge fields. This can be seen from equations (\ref{Leg_transform}) and
(\ref{L_eff-F}).  From the original Einstein equations, we see that we
have to supplement the effective action with a set of constraints.

The result can be split up. In the static case, we have established the
connection with the black hole potential, a concept which we can now
extend to non-spherically symmetric set-ups. The constraints reduce to
one equation, given in (\ref{IndepEinstein}). Also we saw that the
Legendre transform reduces to the entropy function in the literature,
after eliminating the vector fields from the effective action. In the
non-static case, we have found the same electric/magnetic duality
invariant action of \cite{Denef:2000nb}. The new results we obtain are
twofold. On the one hand, we have the new feature of seeing the effective
action as a Legendre transform. On top of that we have now found the
constraints (\ref{multi-constraint2}), stemming from the original
Einstein equations. They have to be supplied to the effective action when
one considers non-supersymmetric setups and were not calculated before.

We hope that this analysis, in particular the explicit form of the
constraints for the non-static effective action, can be a helpful step
towards  (not necessarily supersymmetric) stationary black hole
solutions. Indeed, these constraint are automatically fulfilled for
supersymmetric solutions, see for instance \cite{LopesCardoso:2000qm},
but have to be taken into account when one wants to solve the system in
absence of supersymmetry.

\medskip
\section*{Acknowledgments.}

\noindent We are grateful to Jan De Rydt, Frederik Denef, Renata Kallosh
and Dieter Van den Bleeken for interesting and very useful discussions.
We would also like to thank Oscar Maci{\'a}, who contributed to the early
stages of this work.

This work is supported in part by the European Community's Human
Potential Programme under contract MRTN-CT-2004-005104 `Constituents,
fundamental forces and symmetries of the universe', by the Federal Office
for Scientific, Technical and Cultural Affairs through the
`Interuniversity Attraction Poles Programme -- Belgian Science Policy'
P6/11-P, and in part by the FWO - Vlaanderen, project G.0235.05.  B.V. is
Aspirant FWO-Vlaanderen.

\newpage
\providecommand{\href}[2]{#2}\begingroup\raggedright\endgroup
\end{document}